\documentstyle[aps,epsf]{revtex}  

\epsfclipon

\newcommand{\etc}{{\it etc.}}
\newcommand{\etal}{{\it et al.}}

\newcommand{\vs}{{\it vs.}}

\newcommand{\twiddle}{{$\tilde{\phantom{a}}$}}

\newcommand{\ee}{$e^+e^-$\ }
\newcommand{\mumu}{$\mu^+\mu^-$}
\newcommand{\ave}[1]{\langle#1\rangle}
\newcommand{\simge}{\mathrel{%
   \rlap{\raise 0.511ex \hbox{$>$}}{\lower 0.511ex \hbox{$\sim$}}}}
\newcommand{\simle}{\mathrel{
   \rlap{\raise 0.511ex \hbox{$<$}}{\lower 0.511ex \hbox{$\sim$}}}}
\newcommand{\lsim}{\mathrel {\vcenter {\baselineskip 0pt \kern 0pt
    \hbox{$<$} \kern 0pt \hbox{$\sim$} }}}
\newcommand{\gsim}{\mathrel {\vcenter {\baselineskip 0pt \kern 0pt
    \hbox{$>$} \kern 0pt \hbox{$\sim$} }}}

\begin{document}        
\baselineskip 14pt

\title{
\begin{flushright}
{\small Princeton/$\mu\mu$/99-18}\\
\end{flushright}
Physics Opportunities at a Muon Collider}
\author{Kirk T.~McDonald}
\address{Joseph Henry Laboratories, Princeton University, Princeton, NJ 08544\\
mcdonald@puphep.princeton.edu, http://puhep1.princeton.edu/mumu/}
\maketitle              

\begin{center}   
{\small The case for a future high-energy collider based on muon beams is 
briefly reviewed.}
\end{center}

\section{I Want to Believe...}

\begin{itemize}
\item
That elementary particle physics will prosper for a 2nd century with
laboratory experiments based on innovative particle sources.
\item
That a full range of new phenomena will be investigated:
\begin{itemize}
\item
mass $\Rightarrow$ a 2nd $3 \times 3$ (or larger?) mixing matrix.
\item
Precision studies of Higgs bosons.
\item
A rich supersymmetric sector.
\item
... And more ...
\end{itemize}

\item
That our investment in future accelerators will result in more
cost-effective technology, capable of extension to 10's of TeV
of constituent CoM energy.

\item
That a {\bf Muon Collider} \cite{status,collabpage} based on ionization cooling 
is the best option to accomplish the above.

\end{itemize}

\section{Ionization Cooling} 
\centerline{(An Idea So Simple It Might Just Work)}

\begin{itemize}
\item
Ionization: takes momentum away.
\item
RF acceleration: puts momentum back along $z$ axis.
\item
$\Rightarrow$ Transverse ``cooling''.  

\centerline{\epsfxsize 3.5 truein \epsfbox{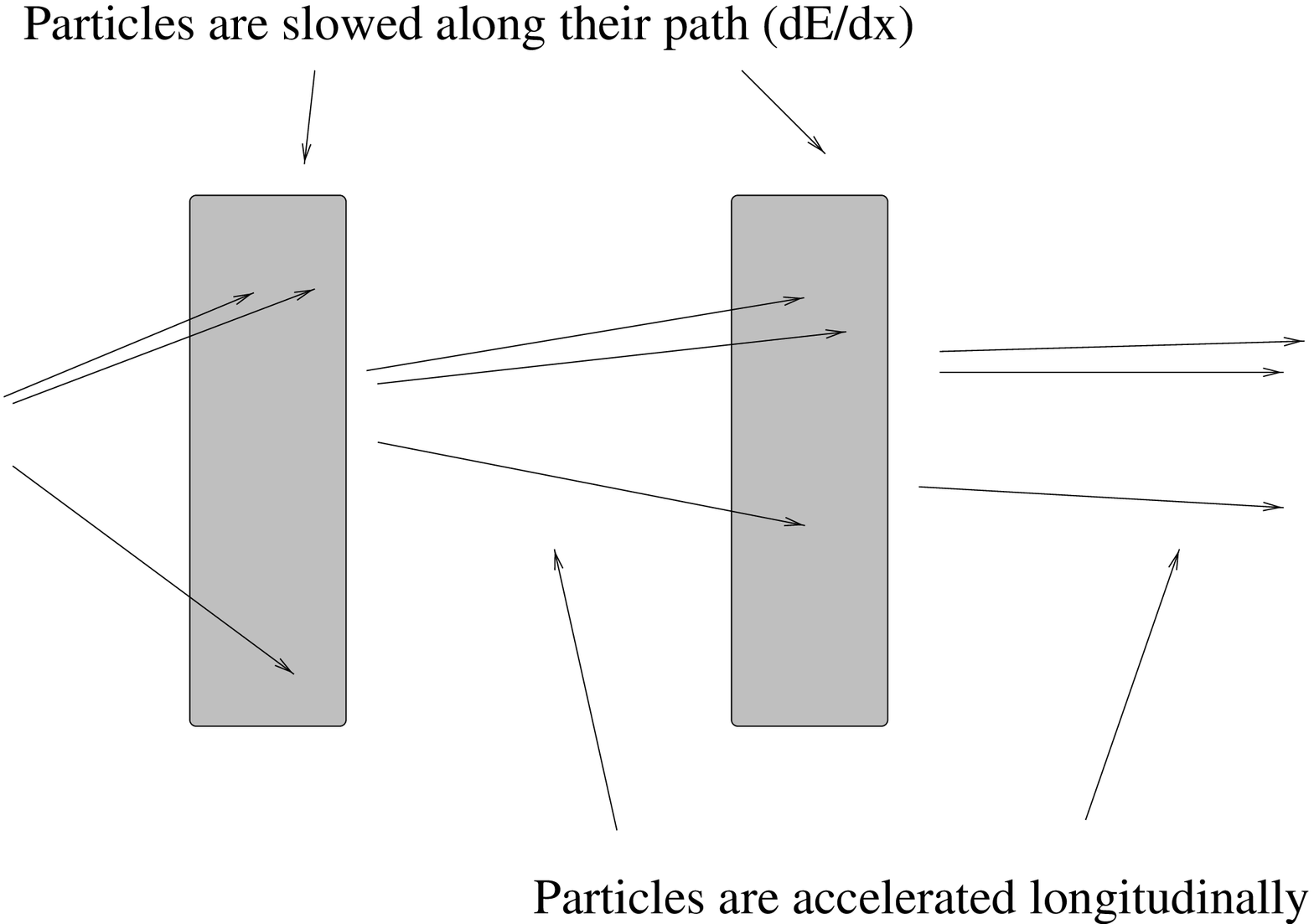}}   

Origin: G.K. O'Neill (1956) \cite{Oneill}.
\item
This  won't work for electrons or protons.
\item
So use muons: Balbekov \cite{Ado}, Budker \cite{Budker}, Skrinsky
\cite{Skrinsky71}, late 1960's.

\end{itemize}

\section{The Details are Delicate}

Use  channel of LH$_2$ absorbers, rf cavities and alternating solenoids
(to avoid buildup of angular momentum).

One cell of the cooling channel:

\centerline{\epsfxsize 3.5 truein \epsfbox{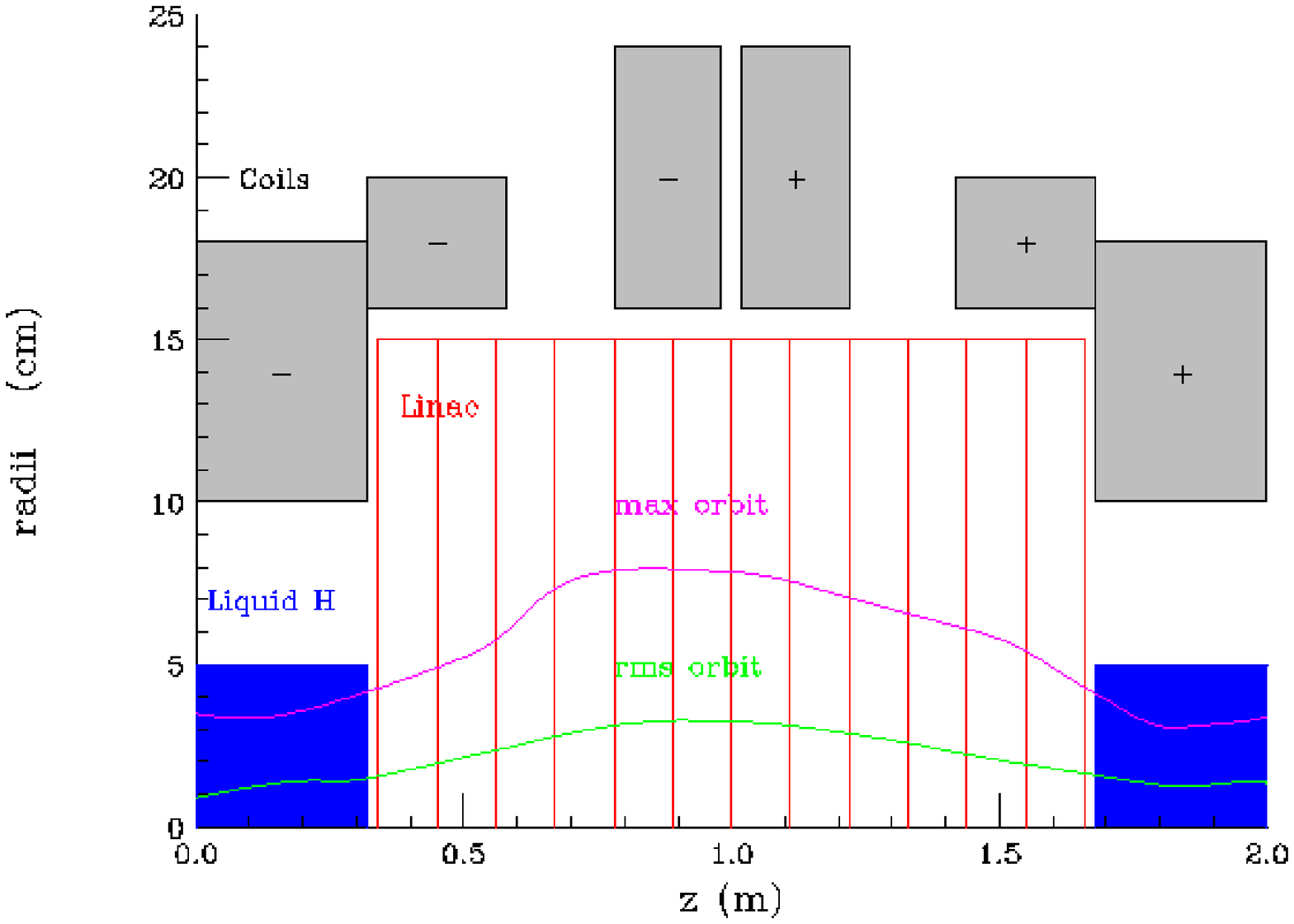}}

But, the energy spread rises due to ``straggling''.

$\Rightarrow$ Must exchange longitudinal and transverse emittance 
frequently to avoid beam loss due to bunch spreading.

Can reduce energy spread by a wedge absorber at a momentum dispersion
point:

\centerline{\epsfxsize 4.0 truein \epsfbox{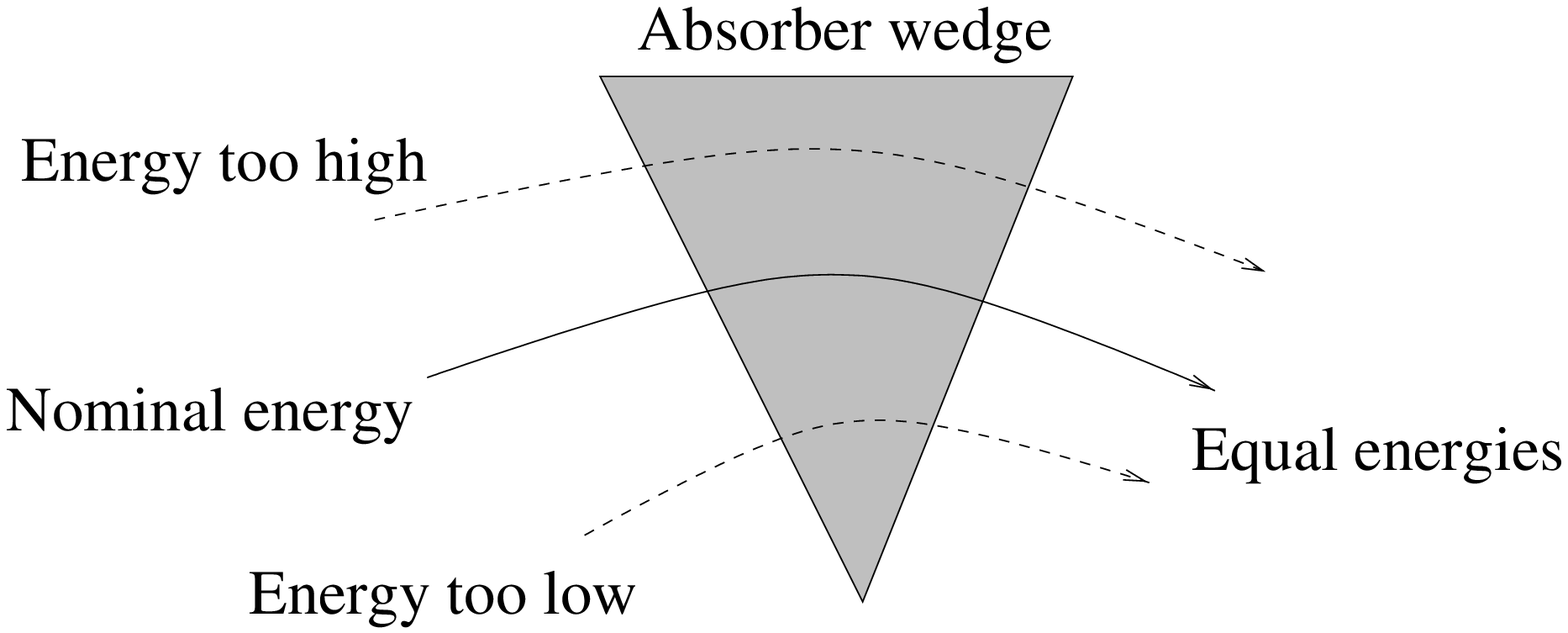}}   

[6-D emittance constant (at best) in this process.]

\section{What is a Muon Collider?}

An accelerator complex in which
\begin{itemize}
\item
Muons (both $\mu^+$ and $\mu^-$) are collected from pion decay 
following a $pN$ interaction.
\item
Muon phase volume is reduced by $10^6$ by ionization cooling.
\item
The cooled muons are accelerated and then stored in a ring.
\item
$\mu^+\mu^-$ collisions are observed over the useful muon life of 
$\approx 1000$ turns at any energy.
\item
Intense neutrino beams and spallation neutron beams are \hfill\break
available as byproducts.
\end{itemize}

Muons decay: $\mu \to e \nu \qquad \Rightarrow$
\begin{itemize}
\item
Must cool muons quickly (stochastic cooling won't do).
\item
Detector backgrounds at LHC level.
\item
Potential personnel hazard from $\nu$ interactions.
\end{itemize}


\newpage

\begin{table}
\caption 
{Baseline parameters for high- and low-energy muon colliders.
Higgs/year assumes a cross section $\sigma=5\times 10^4$~fb; a Higgs width 
$\Gamma=2.7$~MeV; 1~year = $10^7$~s.}
\begin{tabular}{llccccc}
CoM energy         & TeV   & 3 & 0.4 &
\multicolumn{3}{c}{0.1 }  \\
$p$ energy       & GeV        &  16  & 16 & \multicolumn{3}{c}{16}\\
$p$'s/bunch      &    &  $2.5\times 10^{13}$  & $2.5\times 10^{13}$  &
\multicolumn{3}{c}{$5\times 10^{13}$  }  \\  
Bunches/fill   &           & 4 & 4 & \multicolumn{3}{c}{2 }  \\
Rep.~rate  & Hz     &  15 & 15 & \multicolumn{3}{c}{15 }  \\
$p$ power        & MW         &  4   & 4 & \multicolumn{3}{c}{4}  \\ 
$\mu$/bunch  &    & $2\times 10^{12}$ & $2\times 10^{12}$ &
\multicolumn{3}{c}{$4\times 10^{12}$ }  \\
$\mu$ power  & MW     & 28 & 4 & \multicolumn{3}{c}{1 }  \\
Wall power    & MW    &  204 & 120  & \multicolumn{3}{c}{81 }  \\
Collider circum.   & m          &  6000 & 1000 & \multicolumn{3}{c}{350 }  \\
Ave bending field & T       & 5.2 & 4.7 &\multicolumn{3}{c}{3 }  \\
Depth   & m          &  500 & 100 & \multicolumn{3}{c}{10 }  \\
Rms ${\Delta P/P}$       & \%          & 0.16 & 0.14 & 0.12 & 0.01 & 0.003 \\
6d $\epsilon_6$    &  $(\pi \textrm{m})^3$&$1.7\times 10^{-10}$&$1.7\times
10^{-10}$&$1.7\times 10^{-10}$&$1.7\times 10^{-10}$&$1.7\times 10^{-10}$\\
Rms $\epsilon_n$     &$\pi$ mm-mrad     &  50 & 50 & 85 & 195 & 290\\
$\beta^*$         & cm          & 0.3 & 2.6 & 4.1 &  9.4 & 14.1\\
$\sigma_z$         & cm          & 0.3 & 2.6 & 4.1 &  9.4 & 14.1 \\
$\sigma_r$ spot    &$\mu$m     & 3.2 & 26 & 86 & 196 & 294\\
$\sigma_{\theta}$ IP    &mrad     & 1.1 & 1.0 & 2.1 & 2.1 & 2.1\\
Tune shift     &             &0.044 &0.044 & 0.051 &0.022 & 0.015\\
$n_{\rm turns}$ (effective) &     &  785 & 700 & 450 & 450 & 450 \\
Luminosity     & cm$^{-2}$s$^{-1}$ & $7\times 10^{34}$ & $10^{33}$ &
$1.2\times 10^{32}$ & $2.2\times 10^{31}$ & $10^{31}$ \\
 & & & & & & \\
Higgs/year    &  &  & & $1.9\times 10^3$ & $4\times 10^3$ & $3.9\times 10^3$ \\
\end{tabular}
\end{table}


Comparison of footprints of various future colliders:

\centerline{\epsfxsize 6.5 truein \epsfbox{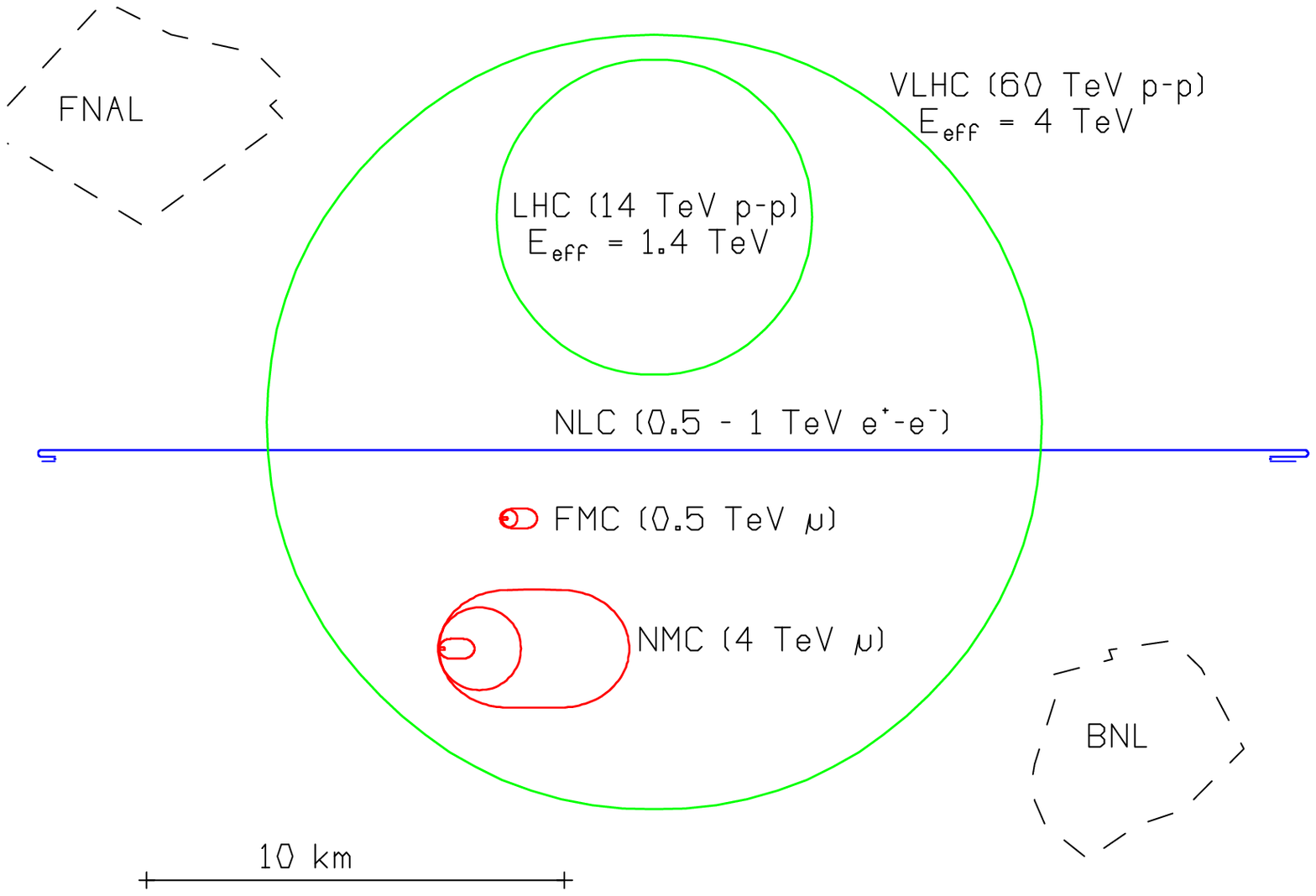}}   

\newpage

A First Muon Collider to study light-Higgs production:

\centerline{\epsfxsize 6.0 truein \epsfbox{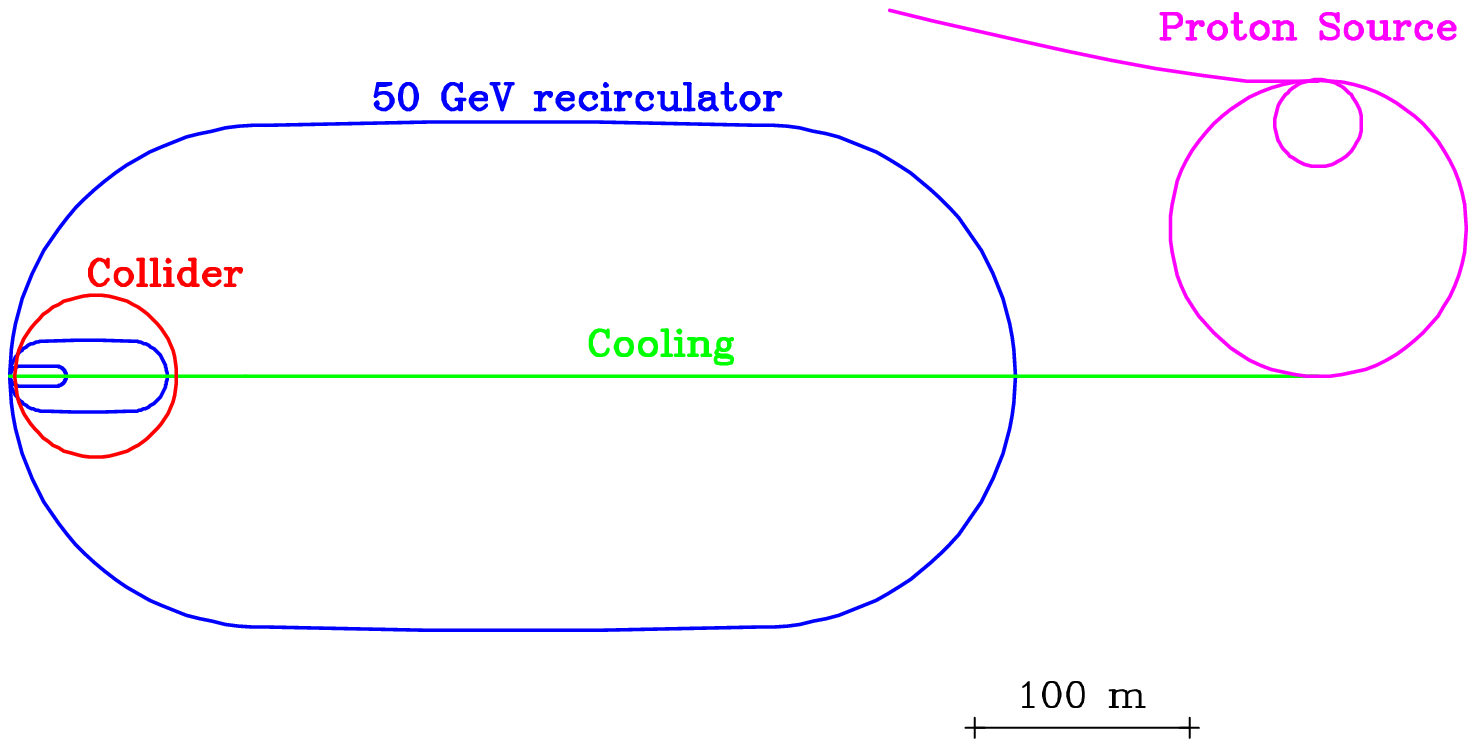}}   

\section{The Case for a Muon Collider}

\begin{itemize}
\item
More affordable than an \ee\ collider at the TeV (LHC) scale.
\item
More affordable than either a hadron or an \ee\ collider for (effective)
energies beyond the LHC.
\item
Precision initial state superior even to \ee.
\hfill\break \phantom{a}\hspace{0.5in}
Muon polarization $\approx 25\%,\ \Rightarrow$ can determine $E_{beam}$ to
$10^{-5}$ via $g-2$ spin precession \cite{Raja}.

\noindent
\parbox{2.5in}
{$t \overline t$ threshold:

{\epsfxsize 2.5 truein \epsfbox{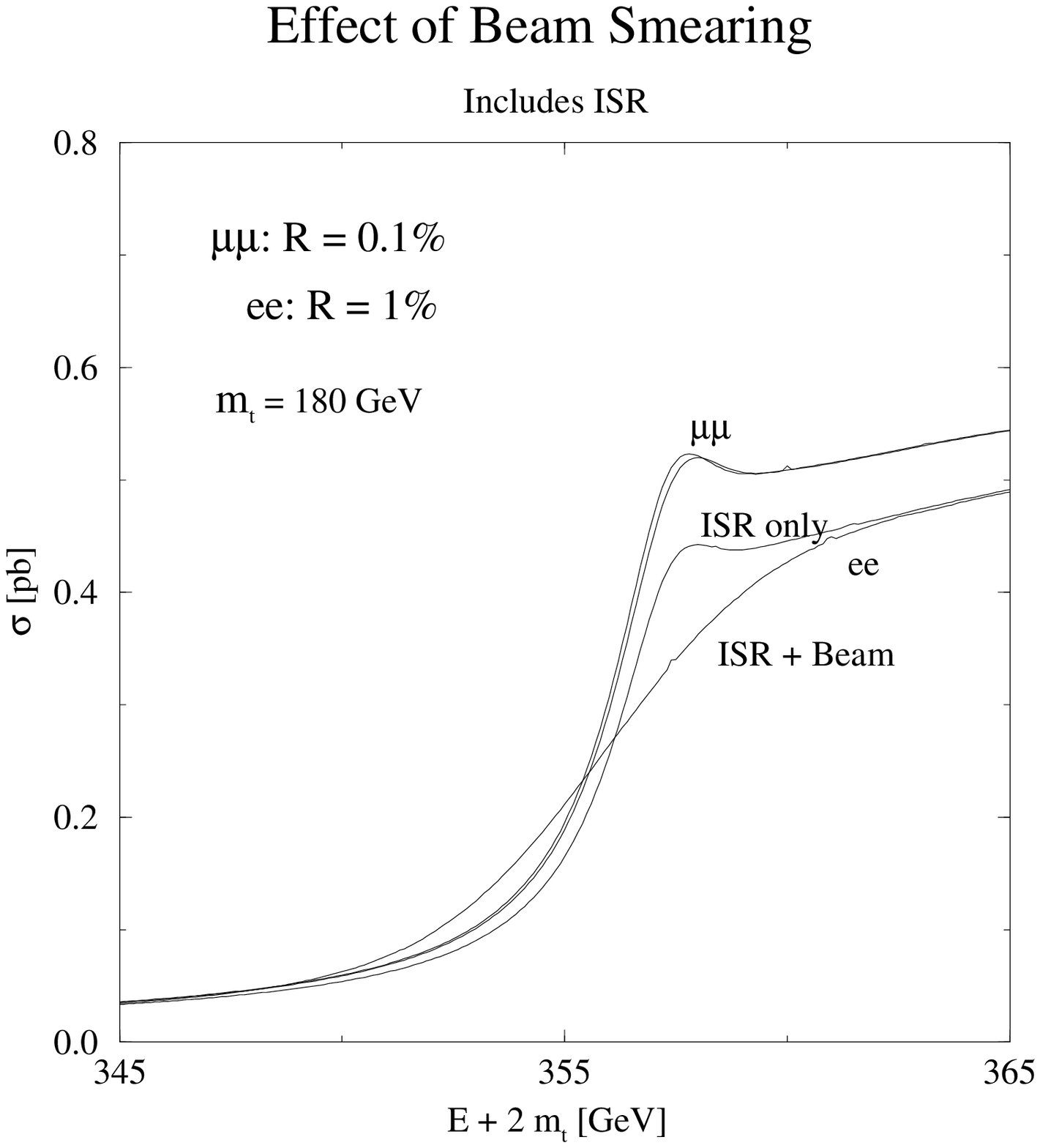}}}
\hspace{0.25in}
\parbox{2.5in}
{Nearly degenerate $A^0$ and $H^0$:

{\epsfxsize 3 truein \epsfbox{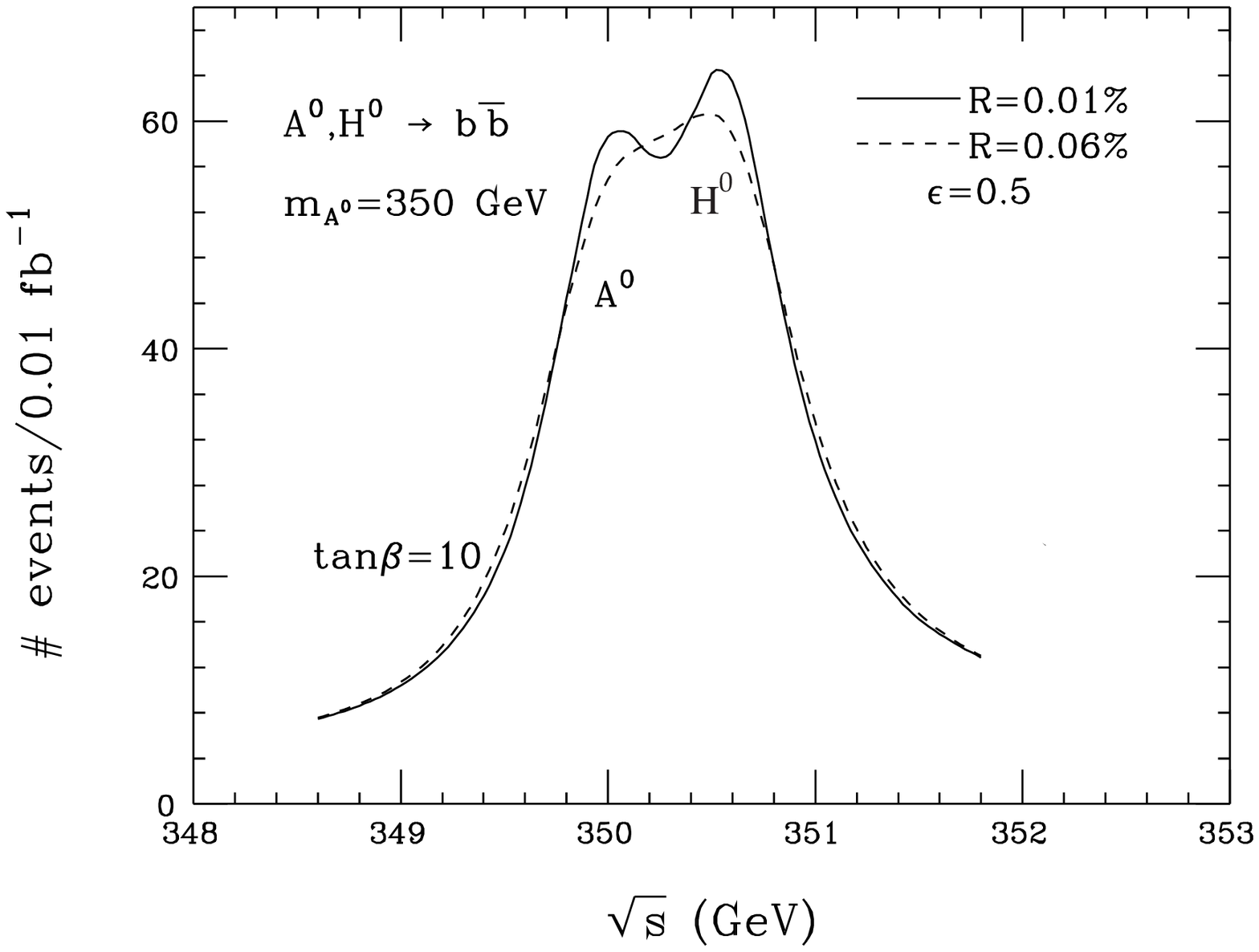}}   }

\item
Initial machine could produce light Higgs via $s$-channel \cite{mup}:
\hfill\break \phantom{a}\hspace{0.5in}
Higgs coupling to $\mu$ is $(m_\mu/m_e)^2 \approx 40,000 \times$ that to 
$e$.
\hfill\break \phantom{a}\hspace{0.5in}
Beam energy resolution at a muon collider $< 10^{-5}$,
\hfill\break \phantom{a}\hspace{1.0in}
$\Rightarrow$ Measure Higgs width.
\hfill\break \phantom{a}\hspace{0.5in}
Add rings to 3 TeV later.

\item
Neutrino beams from $\mu$ decay about $10^4$ hotter than present.
\hfill\break \phantom{a}\hspace{0.5in}
Possible initial scenario in a low-energy muon storage ring \cite{ring}.
\hfill\break \phantom{a}\hspace{0.5in}
$$
\mbox{Study}\ CP\ \mbox{violation via}\ CP\ \mbox{conjugate initial states:}
\left\{ \begin{array}{c} \mu^+ \to e^+ \overline \nu_\mu \nu_e \\
                     \mu^- \to e^- \nu_\mu \overline \nu_e \end{array}
\right. . 
$$


\end{itemize}

\section{Future Frontier Facilities}
\centerline{(A Personal Assessment)}

\begin{itemize}
\item
Hadron collider (LHC, SSC): $\approx$ \$100k/m [magnets].
\hfill\break \phantom{a}\hspace{0.5in}
$\approx$ 2 km per TeV of CM energy.
\hfill\break \phantom{a}\hspace{0.5in}
Ex: LHC has 14-TeV CM energy, 27 km ring, $\approx$ \$3B.
\item
Linear \ee\ collider (SLAC, NLC(?)): $\approx$ \$200k/m [rf].
\hfill\break \phantom{a}\hspace{0.5in}
$\approx$ 20 km per TeV of CM energy; 
\hfill\break \phantom{a}\hspace{0.5in}
But a lepton collider needs only $\approx$ 1/10 the CM energy 
\hfill\break \phantom{a}\hspace{0.5in}
to have equivalent physics reach to a hadron collider.
\hfill\break \phantom{a}\hspace{0.5in}
Ex: NLC, 1.5-TeV CM energy, 30 km long, $\approx$ \$6B (?).
\item
Muon collider: $\approx$ \$1B for source/cooler + \$100k/m for rings
\hfill\break \phantom{a}\hspace{0.5in}
Well-defined leptonic initial state.
\hfil\break \phantom{a}\hspace{0.5in}
$m_\mu/m_e \approx 200 \Rightarrow$ Little beam radiation. 
\hfill\break \phantom{a}\hspace{1.5in} 
$\Rightarrow$ Can use storage rings. 
\hfill\break \phantom{a}\hspace{1.5in} 
$\Rightarrow$ Smaller footprint.
\hfill\break \phantom{a}\hspace{0.5in}
Technology: closer to hadron colliders.
\hfill\break \phantom{a}\hspace{0.5in}
$\approx$ 6 km of ring per TeV of CM energy.
\hfill\break \phantom{a}\hspace{0.5in}
Ex: 3-TeV muon collider, $\approx$ \$3B (?), would have physics reach well
beyond the LHC.

\end{itemize}

\section{Muon Collider R\&D Program}

\begin{itemize}
\item
Targetry and Capture at a Muon Collider Source \cite{targetprop,targetpage}.

Baseline scenario:

\centerline{\epsfxsize 5.0 truein \epsfbox{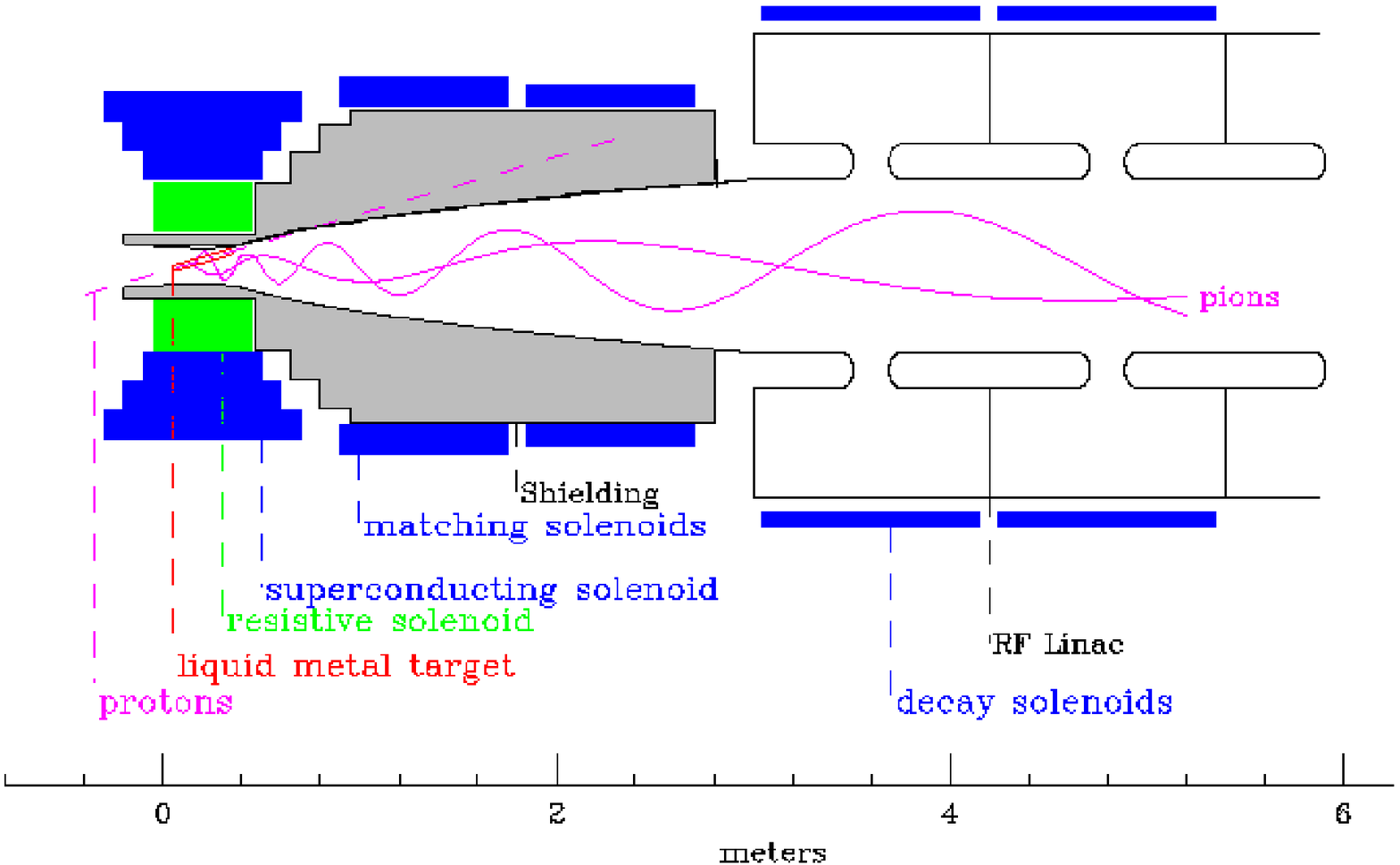}}

To achieve useful physics luminosity, a muon collider must produce about
$10^{14}\ \mu$/sec.
\begin{itemize}
\item
$\Rightarrow > 10^{15}$ proton/sec onto a high-$Z$ target 
$\Leftrightarrow $ 4 MW beam power.
\item
Capture pions of $P_\perp \lsim 200$ MeV/$c$ in
a 20-T solenoid magnet.
\item
Transfer the pions into a 1.25-T-solenoid decay channel.
\item
Compress $\pi/\mu$ bunch energy with rf cavities and deliver to muon 
cooling channel.
\end{itemize}

Proposed R\&D facility:

\centerline{\epsfxsize 5.0 truein \epsfbox{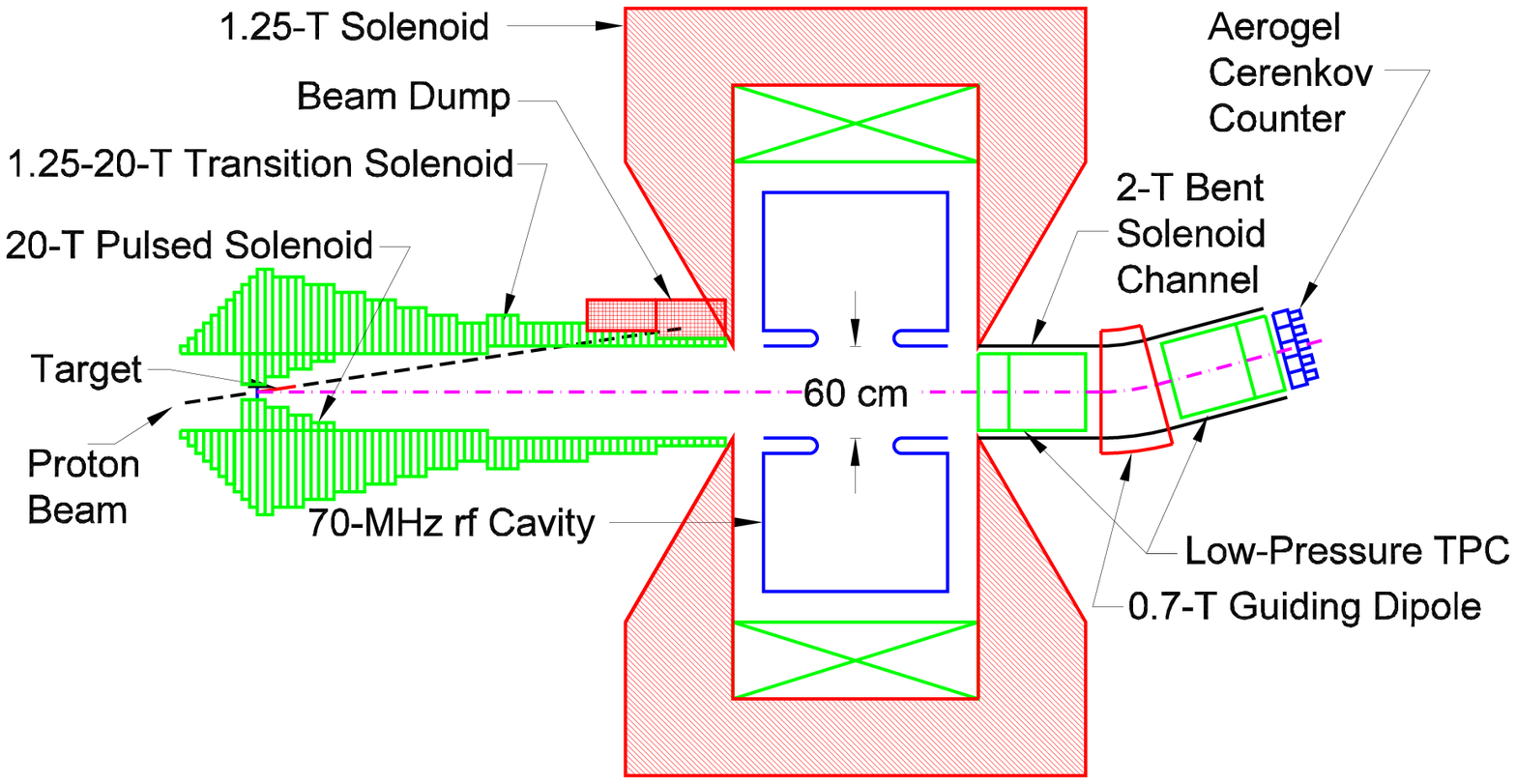}}

\item
Ionization Cooling for a High Luminosity Muon Collider \cite{mucool,coolpage}.

Test basic cooling components:
\begin{itemize}
\item
Alternating solenoid lattice, RF cavities, LH$_2$ absorber.
\item
Lithium lens (for final cooling).
\item
Dispersion + wedge absorbers to exchange longitudinal and
transverse phase space.
\end{itemize}
Track individual muons; simulate a bunch in software.

Possible site: Meson Lab at Fermilab:

\centerline{\epsfxsize 4.5 truein 
            \epsfbox[0 240 490 580]{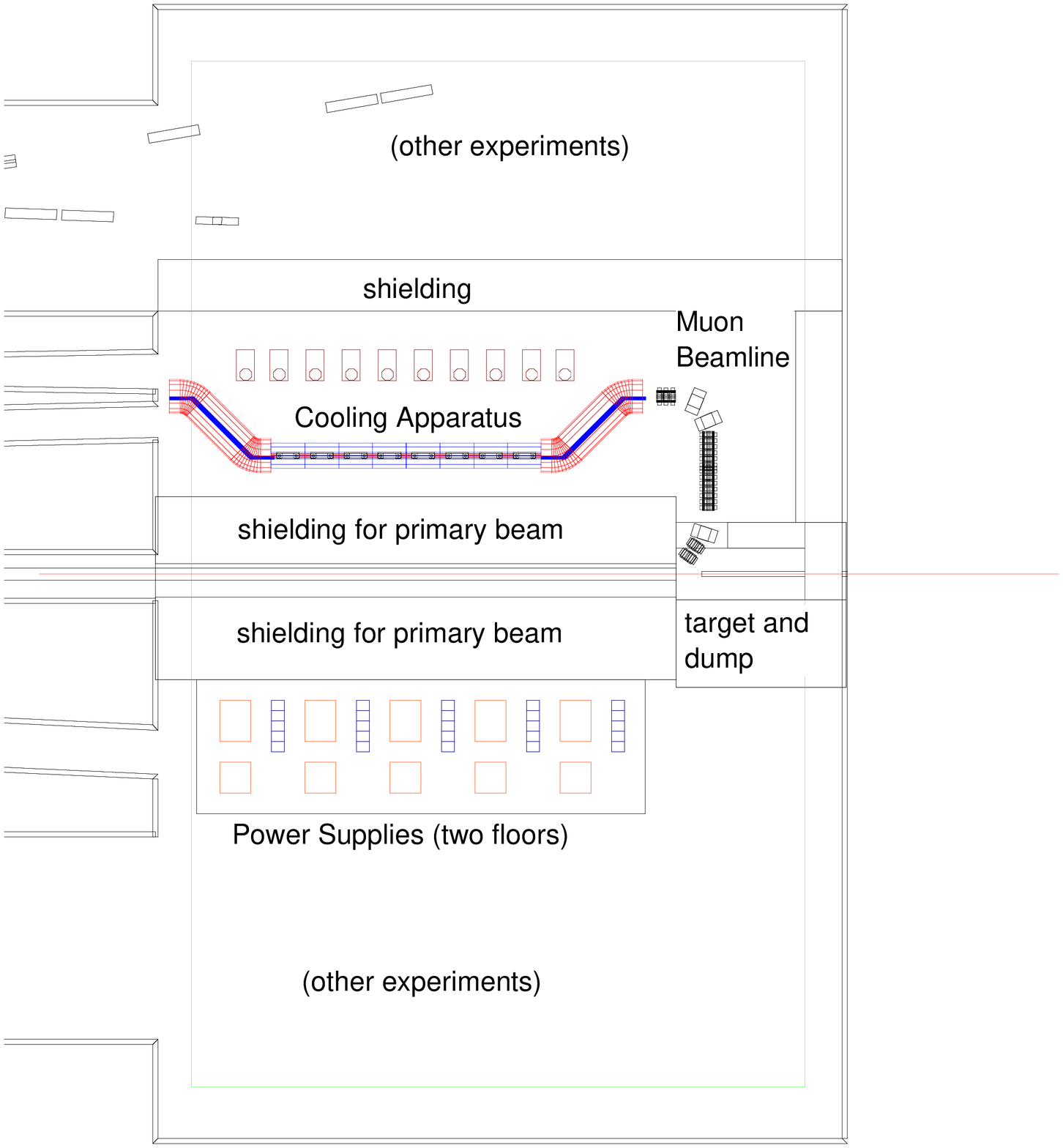}} 

\noindent
\parbox{3.75in}
{Cooling channel components:

\centerline{\epsfxsize 3.75 truein 
            \epsfbox[60 250 450 450]{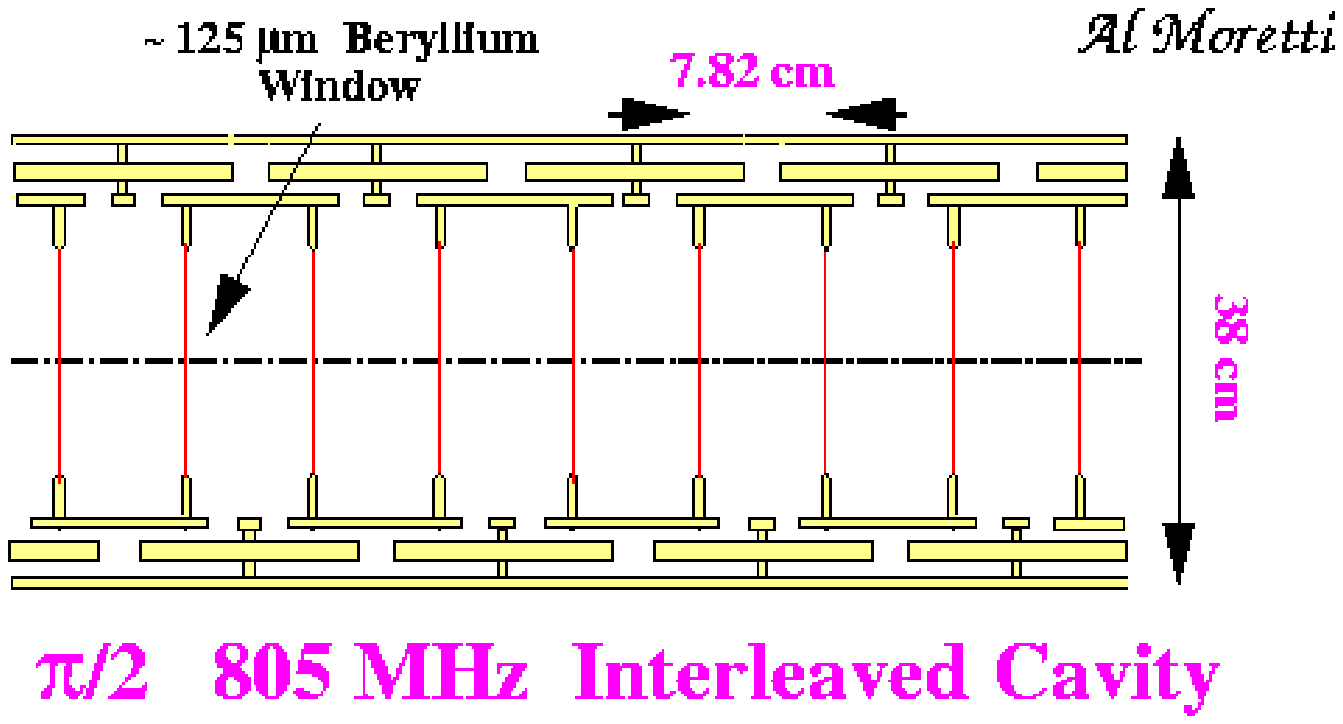}} }
\parbox{2.75in}
{\centerline{\epsfxsize 2.75 truein 
             \epsfbox[90 145 430 410]{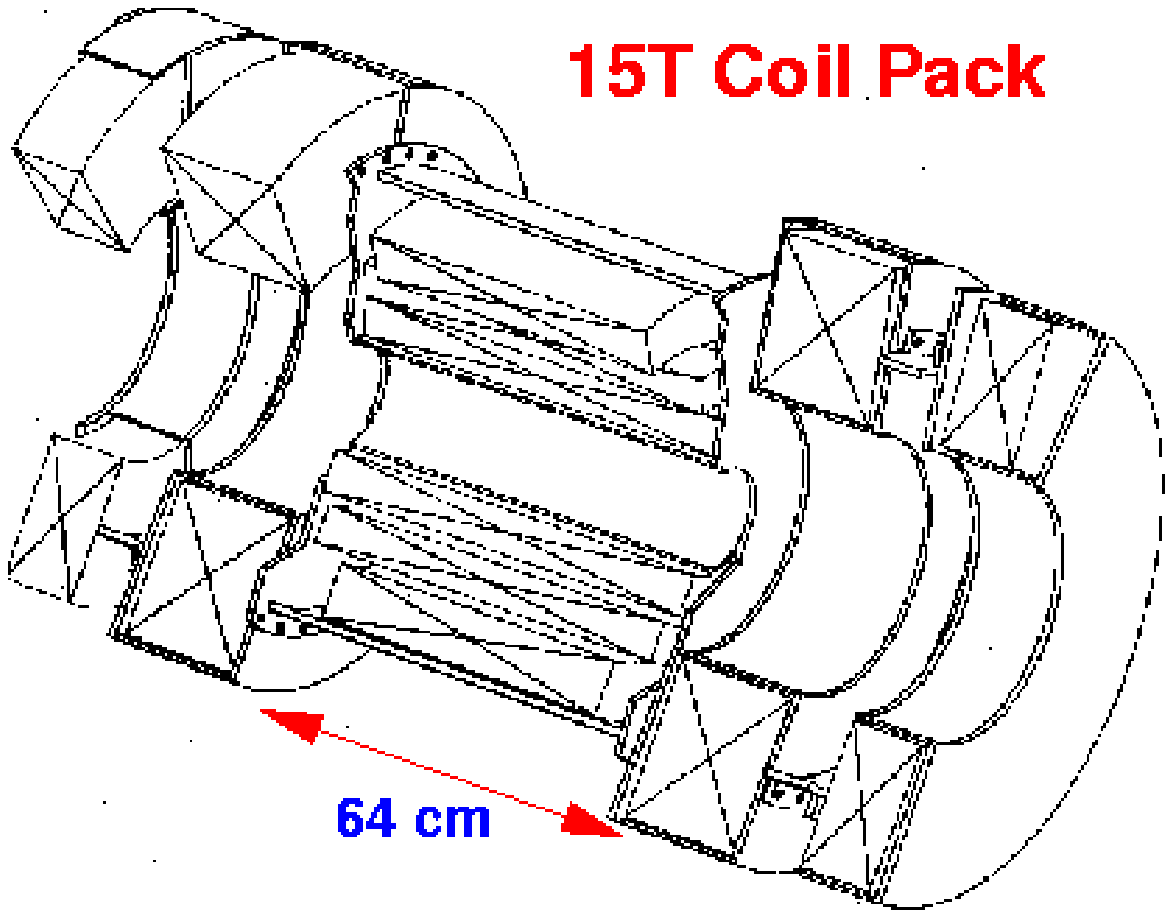}} }

\end{itemize}

\section{Upcoming Workshops}

\centerline{(See http://www.cap.bnl.gov/mumu/table\_workshop.html)}

\begin{itemize}
\item
Muon Collider Collaboration Meeting, May 20-26, 1999, St.~Croix.
\item
Neutrino Factories Based on Muon Accumulators, July 5-9, 1999, Lyon/CERN.
\item
Muon Colliders at the Highest Energies, Sept.\ 27-Oct.\ 1, 1999, Montauk, NY.
\item
Physics Potential and Development of \mumu\ Colliders, Dec.\ 14-19, 1999,
San Francisco.

\end{itemize}


\end{document}